\documentclass [a4paper,fleqn, 12pt]{article}
\usepackage{graphicx}
\usepackage[small]{subfigure,epsfig}

\usepackage {amsmath} \usepackage{amssymb} \usepackage{cite}

\newcommand{\cn}

\begin{document}


\title
{Meromorphic solutions of nonlinear ordinary differential equations}

\author
{Nikolay A. Kudryashov}

\date{Department of Applied Mathematics, National Research Nuclear University
MEPHI, 31 Kashirskoe Shosse,
115409 Moscow, Russian Federation}




\maketitle

\begin{abstract}
Exact solutions of some popular nonlinear ordinary differential equations are analyzed taking their Laurent series into account. Using the Laurent series for solutions of nonlinear ordinary differential equations we discuss the nature of many methods for finding exact solutions. We show that most of these methods are conceptually identical to one another and they allow us to have only the same solutions of nonlinear ordinary differential equations.

\end{abstract}






\section{Introduction}

In the last few years the trend has been towards many publications of exact solutions for nonlinear differential equations. As this phenomenon took place authors told about new solutions of nonlinear partial differential equations but in fact they often find exact solutions of the well - known nonlinear ordinary differential equations. We have numerous examples of papers in which some authors present a lot of exact solutions for nonlinear ordinary differential equations not recognizing that all their solutions are partial cases of the well - known general solution of the equation. Some of these papers were analyzed in the recent papers \cite{ Kudr_2009a, Kudr_2009b, Kudr_2009c, Kudr_2009d, Kudr_2009e, Kudr_2009f, Parkes}. However, publications of 'new solutions' for the well - known differential equations are being continued. Let us give a list of examples.

Salas, Gomez and Hernandez \cite{Salas} presented "new abundant solutions for the Burgers equation". However the authors have used  the traveling wave and studied the nonlinear ordinary differential equation of the first order that is the famous Riccati equation. In essence they have obtained 70 exact solutions of the Riccati equation by means of the Exp - function method.

Li and Zhang  \cite{Li} presented 24 exact solutions of the (2+1) - dimensional Burgers equation. In fact they considered nonlinear ordinary differential equation again taking the traveling wave into account and found 24 exact solutions of the Riccati equation.

Dai and Wang  \cite{Dai_2009} considered the (3+1)-dimensional Burgers
system. However the authors analyzed the Riccati equation as well and found 5 exact solutions of this equation. The authors suppose that some of these solutions are new. We checked all  solutions by Dai and Wang  \cite{Dai_2009} and realized that 3 solutions coincide with the general solution of the Riccati equation and the other 2 expressions do not satisfy this equation.

Wassan \cite{Wazzan} obtained 22 solitary waves solutions of the Korteweg - de Vries equation but in fact he used the traveling wave and considered  nonlinear ordinary differential equation with the well known general solution. As this takes place he believed that 14 solutions are new but all his solutions can be found from well known solutions.

Wen and L\"{u} \cite{Wen_09} introduced "extended Jacobi elliptic function expansion method" and obtained 82 travelling wave solutions of the KdV equation. It is clear that all these exact solutions can be obtained from the well - known general solution.

Wang and Wei \cite{Wang_2010} looked for new exact solutions to the (2+1)-dimensional Konopelchenko-Dubrovsky equation but the authors did find 24 exact solutions of the well-known nonlinear ordinary differential equation.

Aslan \cite{Aslan_2009} "employed the Exp-function method to the Zakharov - Kuznettsov equation as a (2+1) - dimensional model for nonlinear Rossby  waves". In fact the author applied the Exp - function method to the well- known nonlinear ordinary differential equation with the well - known general solution and could not obtain any new results.

Li and Dai \cite{Li_2009} studied the (3+1)-dimensional Jimbo-Miwa equation. The authors obtained 27 exact solutions of well - known nonlinear ordinary differential equation. They studied equation that was solved more than one century ago.

Shang \cite{Shang_2009} "employed the general integral method for traveling wave solutions of coupled nonlinear Klein - Gordon equations". Using the results by Zhang \cite{Zhang_2009} this author presented 36 exact solutions of this system of equations but this system can be reduced to the nonlinear ordinary differential equation with the well - known general solution.

Zheng and Shan \cite{Zheng_2009} applied the Exp - function method to the Whitham-Broer-Kaup shallow water model. These authors believe that they "derived several kinds of new solitary wave solutions". However they presented 8 exact solutions of nonlinear ordinary differential equation with the widely - known general solution.

Yu and Ma \cite{Yu_2009} used the Exp - function method to obtain "the generalized solitary solutions, periodic solutions and other exact solutions for the (2+1) - dimensional KP- BBM wave equation". However, the authors considered the well - known nonlinear ordinary differential equation with the well - known general solution.

Kangalgil and Ayaz F. \cite{Kangalgil_2009} presented 28 exact solutions of the Ostrovsky equation taking the travelling wave into account and using the auxiliary equation method. However it is known that there is a general solution in the form of the travelling wave of the Ostrovsky equation.

Sarma \cite{Sarma} found 4 "exact solutions for modified Korteweg - de Vries equation". Actually, he considered  one of the generalizations of the Korteweg - de Vries equation. This generalization doese not have any exact solution and all expressions by the author do not satisfy this generalization of the KdV equation.

Assas \cite{Assas_2009} obtained 6 "new exact solutions for the Kawahara equation using Exp-function method" but all his solutions do not satisfy the Kawahara equation.

Inan and Ugurlu \cite{Inan} obtained 5 exact solutions of nonlinear evolution equation of the fifth order that is the partial case of the Kawahara equation but all their solutions are wrong.

Zhang with co-authors \cite{Zhang_08} obtained 20 travelling wave solutions for the KdV-Sawada-Kotera equation. However the nonlinear ordinary differential equation corresponding to the KdV - Sawada - Kotera equation was studied many years ago.

We can continue the list of papers with abundant solutions of the well - known nonlinear differential equations more and more. However, we want to know, how many solutions does nonlinear differential equation have?

The aim of this paper is to demonstrate that we can obtain the answer on this question if we look at the Laurent series of solutions for nonlinear ordinary differential equation. We are going to illustrate  as well that the nature of many methods for finding exact solutions of nonlinear differential equations is determined by the Laurent series for solutions of nonlinear differential equations.

Most of the well known nonlinear differential equations have solutions from the class of functions $W$. This class was introduced by Eremenko \cite{Eremenko_2005, Eremenko_2006} and contains meromorphic functions $f(z)$. Function $f(z)$ can be either a rational function of $z$ or a rational function of $\exp{(k\,z)}$ (here $k$ is a complex parameter) or an elliptic function. In fact we can add to this class the Painlev\'e transcendents because a number of solutions of nonlinear differential equations are expressed via the Painlev\'e transcendents.

Let us remind useful information from papers by Eremenko \cite{Eremenko_2005, Eremenko_2006}. Taking the solutions of the Kuramoto - Sivashinsky equation into account Eremenko proved that all known meromorphic solutions of this equation are elliptic functions or their degenerations.   Crucial fact by Eremenko \cite{Eremenko_2005, Eremenko_2006} about solution of the Kuramoto - Sivashinsky equation is the uniqueness property: there is exactly formal meromorphic Laurent series with a movable zero that satisfies the equation.

To check the uniqueness property by Eremenko we need to substitute the series
\begin{equation}\label{Ser1}
w(z)=\sum_{k=m}^{\infty}a_k\, (z-z_0)^k, \quad m<0 \quad a_m\neq0
\end{equation}
into nonlinear ordinary differential equation. However, if we consider the autonomous ordinary differential equation we can use the Laurent series in the form
\begin{equation}\label{Ser2}
w(z)=\sum_{k=m}^{\infty}a_k\, z^k, \quad m<0 \quad a_m\neq0.
\end{equation}
but we have to remember  that we need to change  $z\rightarrow (z-z_0)$  after calculations of coefficients $a_k$.

These statements can be generalized for the most of nonlinear differential equations that are considered by many authors.

This paper is organized as follows. In section 2 we consider the traveling wave solutions of the Burgers equation and point out that these solutions are determined by means of the unique solution of the Riccati equation. We illustrate that there is the unique Laurent series for the solution of the Riccati equation and consequently 'many solutions' of this equation are the same solutions. In section 3 we consider the Korteweg - de Vries equation using the traveling wave solutions as well. We give the Laurent series for traveling wave solutions of this equation. In section 4 we consider the Laurent series for solution of the Ostrovsky equation and obtain that there is the unique solution of this equation. Section 5 is devoted to analysis of the generalization of the Korteweg - de Vries equation. We illustrate that there is no Laurent series for solution of this equation and we can not find any solution of this equation. In section 6 we present the Laurent solution for the travelling wave solution of the Korteweg - de Vries - Burgers equation. We have the Painlev\'e property for this equation as additional condition for parameters of the equation. In this case there is the general solution of this equation that is given in this section. In section 7 we discuss the exact solution of the Kuramoto - Sivashinsky equation and the partial case of this equation. We show that in partial case we obtain only rational solution of this equation. In sections 8 and 9 we discuss the solution of the Kawahara equation and its partial case. We illustrate that the partial case of the Kawahara equation have only rational solution.

\section{Traveling wave solutions of the Burgers equation}

Considering the Burgers equation \cite{Bateman, Burgers, Hopf51, Cole50, Whitham}
\begin{equation}\label{BE1}
u_t+\,u\,u_x=\mu\,u_{xx}
\end{equation}
and using the traveling wave $u(x,t)=y(z)$, where $z=x-C_0\,t$ we obtain the Riccati equation in the form \cite{Salas, Li, Dai_2009}
\begin{equation}\label{BE0}
\mu\,y_{z}-\,\frac{1}{2}\,{y^2}+C_0\,y+C_1=0,
\end{equation}
where $C_1$ is an arbitrary constant of integration.

Taking into account transformations
\begin{equation}\label{BE1}
y(z)={C_0} - 2\mu\,\,w(z)
\end{equation}
we have an equation
\begin{equation}\label{BE2}
w_z=-w^2+\lambda, \qquad \lambda=\,{\frac {{{C_0}}^{2}}{4\,{\mu}^{2}\,{}}}+\,{\frac {{ C_1}}
{2\,{\mu}^{2}{}}}.
\end{equation}

The Laurent series for solution of the Riccati equation takes the form
\begin{equation}\begin{gathered}
\label{Ser3} w_1(z)=\frac{1}{z-z_0}+\frac{\lambda\,(z-z_0)}{3}-\frac{\lambda^2\,(z-z_0)^3}{45}
+\frac{2\,\lambda^3\,(z-z_0)^5}{945}-\\
\\
-\frac{\lambda^4\,(z-z_0)^7}{4725}+
\frac{2\,\lambda^5\,(z-z_0)^9}{93555}+\ldots
\end{gathered}\end{equation}

We can see that there is the unique Laurent series for the solution of the Riccati equation. The Riccati equation is an equation of the first order and the Laurent series contains one  arbitrary constant. Consequently this equation passes the Painlev\'e test. These results were known more than one century ago. We have the general solution of this equation. From the Laurent series \eqref{Ser3} we can see that there is only one solution of the Riccati equation.

We obtain that the traveling wave solutions of the Burgers equation can be found by formula
\eqref{BE1}, where $w(z)$ is a solution of \eqref{BE2} that takes the form
\begin{equation}\begin{gathered}
\label{R2}
w_1(z)=\sqrt{\lambda}\tanh(\sqrt{\lambda}(z-z_0)),\quad \lambda\neq0
\end{gathered}\end{equation}
and
\begin{equation}\begin{gathered}\label{R3}
w_1(z)=\frac{1}{z-z_0},\quad \lambda=0.
\end{gathered}\end{equation}

The general solution of the Riccati equation  is expressed by formula \eqref{R2}, but it is easy to check that the following expressions are exact solutions of Eq.\eqref{BE2} as well

\begin{equation}\begin{gathered}
\label{R5}
w_2(z)=\sqrt{\lambda}\coth(\sqrt{\lambda}(z-c_{2})),\quad \lambda\neq 0
\end{gathered}\end{equation}

\begin{equation}\begin{gathered}
\label{R6}
w_3(z)=\frac{\sqrt{\lambda}}{2}\tanh\left(\frac{\sqrt{\lambda}}{2}(z-c_3)\right)+
\frac{\sqrt{\lambda}}{2}\coth\left(\frac{\sqrt{\lambda}}{2}(z-c_3)\right),\quad \lambda\neq 0
\end{gathered}\end{equation}

\begin{equation}\begin{gathered}
\label{R6a}
w_4(z)=\,\left(\sqrt{\lambda}-\frac{2\,\sqrt{\lambda}}{1+c_4\,
\exp{(2\sqrt{\lambda}\,z})}\right),\quad \lambda\neq 0
\end{gathered}\end{equation}

\begin{equation}\begin{gathered}
\label{R10}
w_5(z)=\frac{c_1\sqrt{\lambda}\exp{(\sqrt{\lambda}z)}-c_5\sqrt{\lambda}\,\exp{(-\sqrt{\lambda}z)}}
{c_1\exp{(\sqrt{\lambda}z)}+c_5\,\exp{(-\sqrt{\lambda}z)}},\quad \lambda\neq 0
\end{gathered}\end{equation}

\begin{equation}\begin{gathered}
\label{R11}
w_6(z)=\frac{c_6\,\sqrt{\lambda}\exp{(\sqrt{\lambda}z)}-\sqrt{\lambda}\,\exp{(-\sqrt{\lambda}z)}}
{c_6\,\exp{(\sqrt{\lambda}z)}+\,\exp{(-\sqrt{\lambda}z)}},\quad \lambda\neq 0
\end{gathered}\end{equation}

\begin{equation}\begin{gathered}
\label{R7}
w_7(z)=-i\,\sqrt{\lambda}\tan(i\,\sqrt{\lambda}(z-c_7)),\quad \lambda\neq 0,\qquad i^2=-1,
\end{gathered}\end{equation}

\begin{equation}\begin{gathered}
\label{R8}
w_8(z)=i\,\sqrt{\lambda}\cot(i\,\sqrt{\lambda}(z-c_8)),\quad \lambda\neq 0,\qquad i^2=-1
\end{gathered}\end{equation}

\begin{equation}\begin{gathered}
\label{R9}
w_9(z)=-\,\frac{i\sqrt{\lambda}}{2}\tan\left(\frac{i\,\sqrt{\lambda}}{2}(z-c_9)\right)+
\frac{i\,\sqrt{\lambda}}{2}\cot\left(\frac{i\,\sqrt{\lambda}}{2}(z-c_9)\right),\quad \lambda\neq 0
\end{gathered}\end{equation}

where $z_0$, $c_1$, $c_2$, \ldots, $c_9$ are arbitrary constants.

The reader can also write other kinds of solutions of the Riccati equation using the identities for trigonometric and hyperbolic functions.

At first glance all solutions $w_1(z)$, $w_2(z)$, \ldots $w_9(z)$ are different. However, it defies common sense to argue that there is the unique Laurent series for the solution of the Riccati equation and have one arbitrary constant $z_0$. How do we understand these facts? The answer is that all these solutions are the same. They are differed by arbitrary constants.  More than that, all forms of the general solution of the Riccati equation $w_1(z)$, $w_2(z)$, \ldots $w_9(z)$ have the same Laurent series \eqref{Ser1}. This Laurent series is changed in case $\alpha=0$ when we have the rational solution of the Riccati equation.

Studying papers \cite{Salas, Li, Dai_2009} we  can see that authors presented a lot of redundant exact solutions of the Riccati equation.

\section{Traveling wave solutions of the KdV equation}

The famous Korteweg - de Vries equation can be written as \cite{Korteweg01}
\begin{equation}\label{KdV1}
u_t+6\,u\,u_x+\,u_{xxx}=0.
\end{equation}
Taking the traveling wave $u(x,t)=w(z)$ into account, where $z=\,x-C_0\,t$ we have nonlinear ordinary differential equation
\begin{equation}\label{KdV0}
\,w_{zz}+\,3\,w^2-C_0\,w+C_1=0.
\end{equation}

The Laurent series for the solution of  equation \eqref{KdV0} takes the form
\begin{equation}\begin{gathered}
\label{Ser3a} w(z)=-\frac{2}{ \left( z-z_{{0}} \right) ^{2}}+\frac{C_{{0}}}{6}
-\left(\frac{C_1}{10}+ \frac{C_0^2}{120}\right) \left( z-z_{{0}} \right)^2+ a_{{6}}\left( z-z_{{0}}\right)^{4}-\\
\\
-{\frac {\left( 12\,C_{{1}}+{C_{{0}}}^{2} \right) ^{2}}{86400}}\, {\left(z-z_0\right)}
^{6}+{\frac {a_{{6}}\left( 12\,C_{{1}}+{C_{{0}}}^{2} \right)}{880}}\, {\left(z-z_0\right)}^{8}+\ldots
\end{gathered}\end{equation}

Eq.\eqref{KdV0} is the second order equation. This equation has parameters $C_0$ and $C_1$. We can see that we have the unique Laurent series for the solution of equation \eqref{KdV0} as well. The Laurent series for solution of Eq.\eqref{KdV0} has two arbitrary constants $z_0$ and $a_6$. Eq.\eqref{KdV0} passes the Painlev\'e test. There is the unique general solution that is expressed via the Jacobi elliptic function in general case. In the case $C_0=C_1=a_6=0$ we have the well known rational solution of the Korteweg - de Vries equation
\begin{equation}\begin{gathered}
\label{Ser3a} w(z)=-\frac{2}{ \left( z-z_{{0}} \right) ^{2}}.
\end{gathered}\end{equation}

Let us present the general solution of Eq. \eqref{KdV0}.
Multiplying Eq. \eqref{KdV0} by $w_{z}$ and integrating this
equation with respect to $z$, we have the nonlinear differential
equation in the form
\begin{equation}\begin{gathered}
\label{KdV.2} w_{z}^{2}+2\,w^3-\,C_0\,w^2 +2\,C_1\,w+2\,C_2=0,
\end{gathered}\end{equation}
where $C_2$ is the second constant of integration.

Assuming that  $\alpha$, $\beta$ and $\gamma$
($\alpha\,\geq\,\beta\,\geq\,\gamma$) are roots of the algebraic
equation
\begin{equation}\begin{gathered}
\label{KdV.3} w^3-\frac12\,C_0\,w^2 +C_1\,w+C_2=0,
\end{gathered}\end{equation} we can write Eq. \eqref{KdV.2} in the form
\begin{equation}\begin{gathered}
\label{KdV.4} w_{z}^{2}=-2\,(w-\alpha)\,(w-\beta)\,(w-\gamma).
\end{gathered}\end{equation}

Comparison of Eq. \eqref{KdV.3} and Eq. \eqref{KdV.4} allows us to
find relations between the roots $\alpha$, $\beta$, $\gamma$ and the
constants $\omega$, $C_1$, $C_2$ in the form
\begin{equation}\begin{gathered}
\label{KdV.6} \alpha\,\beta\,\,\gamma=-{C_2},\quad
\alpha\,\beta+\alpha\,\gamma+\beta\,\gamma=C_1, \quad
\alpha+\beta+\gamma=\frac{C_0}{2}.
\end{gathered}\end{equation}

The general solution of Eq. \eqref{KdV.2} is expressed via the
Jacobi elliptic function \cite{Drazin01, King01}
\begin{equation}\begin{gathered}
\label{KdV.5}
w{(z)}=\beta+(\alpha-\beta)\,cn^2{\left\{\sqrt{\frac{\alpha-\gamma}{2}}\,z,\,\,
S^2\right\}}, \quad S^2=\frac{\alpha-\beta}{\alpha-\gamma},
\end{gathered}\end{equation}
where $cn(z)$ is the elliptic cosine. The general solution of
equation \eqref{KdV.5} was first found by Korteweg and de Vries
\cite{Korteweg01}.

The solitary wave solutions of Eq. \eqref{KdV0} are found as partial cases of solution \eqref{KdV.5} when Eq. \eqref{KdV.3} has two equal roots \cite{Kudr_2009b}. There is another approach in finding solitary wave solutions of Eq.\eqref{KdV0}. Aiming it we can take the solution of the Riccati equation \eqref{BE2}. Assuming
\begin{equation}\begin{gathered}
\label{KdV.6}
w{(z)}=A_0+A_1\,w_1(z)+A_2\,w_1(z)^2
\end{gathered}\end{equation}
and substituting \eqref{KdV.6} into Eq.\eqref{KdV0} we have
\begin{equation}\begin{gathered}
\label{KdV.7}
A_2\,=-2,\qquad A_1=0,\qquad A_0=\frac16\,{C_0}+\frac43\,\lambda,\\
\\
C_1=-\frac43\,{\lambda}^{2}+\frac{1}{12}\,{{C_0}}^{2}\qquad
C_2=\frac49\,{C_0}\,{\lambda}^{2}-{\frac {1}{108}}\,{{ C_0}}^{3}-{\frac
{32}{27}}\,{\lambda}^{3}
\end{gathered}\end{equation}

The well - known solitary wave solution takes the form
\begin{equation}\begin{gathered}
\label{KdV.8}
w(z)=\frac16\,{C_0}+\frac43\,\lambda-2\,\lambda\, \tanh^2 \left( \sqrt {
\lambda}(z-z_0) \right)
\end{gathered}\end{equation}
This soliton solution can be written in different forms but it is a well - known partial case of the general solution \eqref{KdV.5}. Consequently a lot  of exact solutions of the Korteweg - de Vries equation by the authors \cite{Wazzan, Wen_09} are redundant. We can say similar arguments  about exact solutions of nonlinear ordinary differential equations considered in \cite{Wang_2010, Aslan_2009, Li_2009, Shang_2009, Zhang_2009, Zheng_2009,Yu_2009} because the Kopelchenko - Dubrovsky equation, the Jimbo - Miwa equation, the Whitham - Broer -Kaup equation, the Zakharov - Kuznetsov equation and the KP - BMM equation by means of the traveling wave can be reduced to Eq.\eqref{KdV0}.

Using the Laurent series for different forms of solution of the Riccati equation we can understand the nature of many methods for finding exact solutions of nonlinear differential equations. We can see that most of these approaches are the same. Actually we can use any functions  $w_1(z)$, $w_2(z)$, \ldots $w_9(z)$ to search for the exact solutions of nonlinear ordinary differential equations but we have to know that we can obtain only the same solutions. More than that, we can use the Laurent series \eqref{Ser3} for finding exact solutions of nonlinear ordinary differential equations.

 We can see that if we use the solution $w_1(z)$ of the Riccati equation we have the tanh - function method. Taking the solution $w_3(z)$ of the Riccati equation into account we obtain tanh - coth method to look for exact solutions of nonlinear ordinary differential equations. Using the solution $w_5(z)$ we get the $G^{'}/G$ - method.

We can also understand the nature of the Exp - function method taking the solution $w_5(z)$ into account. We do not like the application of the Exp - function method in searching exact solutions of nonlinear differential equations because this method does not tell us anything about the pole order of solution of nonlinear ordinary differential equation. Application of this method for nonlinear ordinary differential equation can give exact solutions by exhaustive search of different sums of exponential functions in nominator and denominator. The Exp - function method did not allow to find exact solutions with the pole of the third and the fourth order through cumbersome calculations. But the application of this method for equations with  solutions of the  first and the second order pole can not give anything new in comparison with other more simple methods.

Let us note that all these approaches are conceptually identical with the simplest equation method which was formulated in papers \cite{Kudr_05, Kudr_05a}. The idea of the simplest equation method is to  find the relation between two equations. One of these equations is the equation with unknown solution. We want to look for exact solution of this equation. Another equation has the well - known solution. We can choose this equation. As a simplest equation we can use the Riccati equation, equation for Jacobi elliptic function, equation for Weierstrass function, equation with the Painlev\'e transcendents and so on. In essence, using the simplest equation method we take the Laurent series for solutions of two equations into account.

\section{Traveling wave solutions of the Ostrovsky equation}

 Consider the Ostrovsky equation \cite{Parkes_07}
\begin{equation}
  \begin{gathered}
  \label{E1}u\,u_{xxt}-u_x\,u_{xt}+u^2\,u_t=0
  \end{gathered}
\end{equation}
Using the traveling wave
\begin{equation}
  \begin{gathered}
  \label{E2}u(x,t)=w(z),\,\qquad z=x-C_0\,t
  \end{gathered}
\end{equation}
K\"{o}roglu and \"{O}zis \cite{Koroglu_2009} tried to "predict a new traveling wave solution" of equation
\begin{equation}
  \begin{gathered}
  \label{E3} \,w\,w_{zzz}-\,w_{z}\,w_{zz}+w^2\,w_{z}=0.
  \end{gathered}
\end{equation}
The authors applied the Exp-function method and have believed that they have found a new exact solution of Eq.\eqref{E3}.

The Laurent series for the general solution of Eq.\eqref{E3} can be written in the form
\begin{equation}
  \begin{gathered}
  \label{E3a} w(z)=-\frac{6}{(z-z_0)^2}+{a_2}-\frac{{{a_2}}^{2}}{10}\,{(z-z_0)}^{2}+{a_6}\,{(z-z_0)}^{4}-\\
  \\
  -{\frac {{{a_2}}^{4}}{1800}}\,{(z-z_0)}^{6}+{\frac {{{a_2}}^{
2}{a_6}}{220}}\,\,{(z-z_0)}^{8}-\left({\frac {{{a_2}}^{6}}{702000}}\,+{
\frac {{{a_6}}^{2}}{78}} \right) {(z-z_0)}^{10}
+\ldots
\end{gathered}\end{equation}

We have three arbitrary constants $z_0$, $a_2$ and $a_6$ in the unique Laurent series \eqref{E3a} for the general solution of Eq.\eqref{E3}. Consequently Eq.\eqref{E3} passes the Painlev\'e test and we can expect to find the general solution of Eq.\eqref{E3}.

In the case $a_2=a_6=0$ from the Laurent series \eqref{E3a} we find the rational solution of Eq.\eqref{E3} in the form
\begin{equation}
  \begin{gathered}
  \label{E3b} w(z)=-\frac{6}{(z-z_0)^2}.
  \end{gathered}
\end{equation}

In general case Eq.\eqref{E3} can be written as
\begin{equation}
  \begin{gathered}
  \label{E4} \,\frac{d}{dz}\left(\frac{1}{w}\frac{d^2w}{dz^2}\right)+\frac{dw}{d{z}}=0.
  \end{gathered}
\end{equation}
Integrating Eq.\eqref{E4} with respect to $z$ we have
\begin{equation}
  \begin{gathered}
  \label{E5} \,w_{zz}+w^2-C_1\,w=0,
  \end{gathered}
\end{equation}
where $C_1$ is a constant of integration.

Multiplying Eq.\eqref{E5} on $w_{z}$ and integrating expression again with respect to $z$  we obtain
\begin{equation}
  \begin{gathered}
  \label{E6}\,w_{z}^{2}+\frac{2}{3}\,w^3-C_1\,w^2\,-C_2=0,
  \end{gathered}
\end{equation}
where $C_2$ is a constant of integration as well.

The theory of ordinary differential equations has more than three hundred years of history and we think that the hope by using the simple methods, one can obtain new results, is naive. However, in two last decades we observe the explosion of papers where the authors present 'new methods for finding exact solutions' and 'new solutions of nonlinear ordinary differential equations'.

Eq.\eqref{E6} is a partial case of the well known ordinary differential equation
\begin{equation}
  \begin{gathered}
  \label{E6a}w_{z}^{2}-4\,w^3-C_1\,w^2\,-C_3\,w-C_2=0,
  \end{gathered}
\end{equation}
where $C_1$, $C_2$ and $C_3$ are arbitrary constants.

Eq. \eqref{E6a} is well known in physics and mathematics.

We can present Eq.\eqref{E6} in the form
\begin{equation}
  \begin{gathered}
  \label{E7} \int \frac{dw}{\sqrt{C_2+C_1\,w^2-\frac{2}{3}\,w^3}}={z-z_0}
  \end{gathered}
\end{equation}

Let $\alpha$, $\beta$ and $\gamma$ ( $\alpha \geq \beta \geq \gamma$) be real roots of the equation
\begin{equation}
  \begin{gathered}
  \label{E8} w^3-\frac{3}{2}\,C_1\,w^2-\frac{3}{2}\,C_2=0
  \end{gathered}
\end{equation}
then the general solution of Eq.\eqref{E7} can be presented in the form
\begin{equation}
  \begin{gathered}
  \label{E9} \int \frac{dw}{\sqrt{(\alpha-w)\,(w-\beta)\,(w-\gamma)}}=\sqrt{\frac23}\,{(z-z_0)}.
  \end{gathered}
\end{equation}
Taking into account in \eqref{E9} the following variables
\begin{equation}
  \begin{gathered}
  \label{E10} w=\alpha-(\alpha-\beta)\,q^2,\qquad q\equiv q(z)
  \end{gathered}
\end{equation}
we  have an expression
\begin{equation}
  \begin{gathered}
  \label{E11}sn^{(-1)}\left\{q;S^2\right\}=\int_{0}^{q} \frac{d y}{\sqrt{(1-y^2)\,(1-S^2\,y^2)}}=\\
  \\
  =\frac{(z-z_0)}{\alpha}\,\sqrt{\frac{\alpha-\gamma}{6}},\,\qquad S^2=\frac{\alpha-\beta}{\alpha-\gamma}
  \end{gathered}
\end{equation}
Taking \eqref{E10} and \eqref{E11} into consideration we have
\begin{equation}
  \begin{gathered}
  \label{E12}U(\xi)=\alpha+(\alpha-\beta)\, sn^2\left\{{(z-z_0)}{}\,\sqrt{\frac{\alpha-\gamma}{6}};\,S^2\right\}.
  \end{gathered}
\end{equation}
Finally we can write the general solution of Eq.\eqref{E3} in the form
\begin{equation}
  \begin{gathered}
  \label{E12}w(z)=\beta+(\alpha-\beta)\, cn^2\left\{{(z-z_0)}{}\,\sqrt{\frac{\alpha-\gamma}{6}};\, S^2\right\}.
  \end{gathered}
\end{equation}

The solitary wave solution can be found from \eqref{E12}. However we can find these solutions  taking into account the Laurent series for solutions of the Riccati equation and
\begin{equation}
  \begin{gathered}
  \label{E13}w(z)=A_0+A_1\,w_1(z)+A_2\,w_1(z).
  \end{gathered}
\end{equation}

Substituting the Laurent series $w_1(z)$  and $w_2(z)$ for the solutions of the Riccati equation and for the Ostrovsky equation we have
\begin{equation}
  \begin{gathered}
  \label{E14}A_2=-6,\quad A_1=0, \quad A_0^{(1)}=2\,\lambda, \quad A_0^{(2)}=6\,\lambda
  \end{gathered}
\end{equation}
As a result we have two solitary wave solutions of the Ostrovsky equation
\begin{equation}
  \begin{gathered}
  \label{E15} w^{(1)}(z)=2\,\lambda-6\,\lambda\,  \tanh^{2} \left( \sqrt {\lambda}z \right)
 ,
  \end{gathered}
\end{equation}
and
  \begin{equation}
  \begin{gathered}
  \label{E16} w^{(2)}(z)=6\,\lambda-\,6\,\lambda\,  \tanh^{2} \left( \sqrt {\lambda}z \right)
 .
  \end{gathered}
\end{equation}

All solitary wave solutions of papers \cite{Kangalgil_2009, Koroglu_2009} can be found from expression Eqs.\eqref{E15} and \eqref{E16}.

Therefore, authors \cite{Kangalgil_2009, Koroglu_2009} have not found any new solutions of the Ostrovsky equation using the Exp-function method.

\section{One of the generalizations of the KdV equation}

Consider one of generalizations of the Korteweg - de Vries equation which can be written in the form \cite{Sarma}
\begin{equation}\label{KdV1a}
u_t+6\,u\,u_x+\,u_{xxx}=\alpha\,u.
\end{equation}

Using the traveling wave, we have a nonlinear differential equation
\begin{equation}\label{KdV1b}
\,w_{zzz}+\,6\,w\,w_z-C_0\,w_z-\alpha\,w=0.
\end{equation}

Substituting the Laurent series \eqref{Ser1} into Eq.\eqref{KdV1b} we get $m=-2$ and  coefficients $a_0$, $a_1$, $a_2$, $a_3$, $a_4$ and $a_5$ in the form
\begin{equation}\begin{gathered}
\label{KdVd} a_0=-2,\quad a_1=0, \quad a_2=\frac{C_{{0}}}{6}, \quad a_3=-\frac{\alpha\,}{6}, \quad a_4 =C_1, \quad a_5=-\frac{\alpha\,{C_0}\,}{36}
\end{gathered}\end{equation}
where $C_1$ is an arbitrary constant. Equation \eqref{KdV1b} is of the third order, but we obtained only two arbitrary constants $z_0$ and $a_4$. We can not find $a_6$ in the Laurent series \eqref{Ser1} for the solution of Eq.\eqref{KdV1b} in case  $\alpha\neq0$ because the Laurent series does not exist.  We do not have the meromorphic solutions of Eq.\eqref{KdV1b}. We expect that exact solutions of Eq.\eqref{KdV1b} can not be found. However Sarma in \cite{Sarma} presented 4 exact solutions of Eq.\eqref{KdV1b}. We have checked "exact solutions" by Sarma \cite{Sarma} and confirmed that all of them do not satisfy Eq.\eqref{KdV1b}. Hence his solutions are wrong.

\section{Traveling wave solutions of the KdV - Burgers equation}

Consider the Korteweg - de Vries - Burgers equation
\begin{equation}\label{KdVB1}
u_t+\,u\,u_x+\,u_{xxx}=\alpha\,u_{xx}.
\end{equation}
Using the traveling wave solutions $u(x,t)=w(z)$, where $z=\,x-C_0\,t$ we have the nonlinear ordinary differential equation of the second order
\begin{equation}\label{KdVB2}
w_{zz}-\alpha\,w_z+\,\frac12\,w^2-C_0\,w+C_1=0.
\end{equation}
where $C_1$ is an arbitrary constant.

The Laurent series for solution of Eq.\eqref{KdVB2} can be found at $C_1=\frac{{{ C_0}}^{2}}{2}\,-{\frac {18\,\alpha^4}{625}}$ (in this case $a_6$ is an arbitrary constant) and takes the form
\begin{equation}\begin{gathered}
\label{KdVB3} w(z)=-\frac{12}{(z-z_0)^2}-\frac{12\,\alpha}{5(z-z_0)}+{C_0}{}+{\frac {\alpha^2}{25}}-{\frac {\alpha^3(z-z_0)}{
125}}\,-\\
\\
-{\frac {\alpha^4(z-z_0)^2}{12500}}\,+
{\frac {{\alpha}^{5}{(z-z_0)}^{3}}{187500}}\,
+{a_6}\,{(z-z_0)}^{4}+\\
\\
+\left( \frac{4\,\alpha\,{a_6}}{5}\,-{\frac {{\alpha}^{7}}{9375000}}\,
 \right) {(z-z_0)}^{5} + \ldots
\end{gathered}\end{equation}

We can see that  Eq.\eqref{KdVB1} has passed the Painlev\'e test in the case  $C_1=\frac{{{ C_0}}^{2}}{2}\,-{\frac {18\,\alpha^4}{625}}$. In this case there are two arbitrary constants $z_0$ and $a_6$ in the Laurent series \eqref{KdVB3}. At arbitrary parameters $\alpha$ and $C_1$ the Laurent series does not exist. In the general case there is no meromorphic solution of  Eq.\eqref{KdVB1}. However when $C_1=\frac{{{ C_0}}^{2}}{2}\,-{\frac {18\,\alpha^4}{625}}$ the general solution of Eq.\eqref{KdVB1} was obtained. Let us give this solution again.

Let us assume
\begin{equation}\begin{gathered}
\label{KdVB.10a} w(z)=b-v(z)
\end{gathered}\end{equation}
where $b$ is a constant which will be found. Substituting \eqref{KdVB.10a}
into \eqref{KdVB2} we obtain equation
\begin{equation}\begin{gathered}
\label{KdVB.5}v_{zz}-\alpha\,v_{z}-\frac12\,v^2+(b-C_0)\,v
+b\,C_0-\frac{b^2}{2}-C_1=0.
\end{gathered}\end{equation}

Assuming in \eqref{KdVB.5}
\begin{equation}\begin{gathered}
\label{KdVB.6}C_1=b\,C_0-\frac{b^2}{2}
\end{gathered}\end{equation}
we get equation
\begin{equation}\begin{gathered}
\label{KdVB.7}\,v_{zz}-{\alpha}{}\,v_{z}-
\frac{1}{2\,}\,v^2+{(b-C_0)}{}\,v=0.
\end{gathered}\end{equation}

Let us search for solutions of Eq.\eqref{KdVB.7} using the new
variable
\begin{equation}\begin{gathered}
\label{KdVB.8}v(z)=e^{-m\,z}\,W(z),
\end{gathered}\end{equation}
where $m$ is an unknown parameter to be found.

Taking \eqref{KdVB.8} into account we have
\begin{equation}\begin{gathered}
\label{KdVB.9}v_{z}=(W_{z}-m\,W)\,e^{-m\,z}, \qquad
v_{zz}=(m^2\,W-2\,m\,W_{z}+W_{zz})\,e^{-m\,z}.
\end{gathered}\end{equation}

Substituting \eqref{KdVB.8} and \eqref{KdVB.9} into Eq.
\eqref{KdVB.7} we obtain the equation
\begin{equation}\begin{gathered}
\label{KdVB.10}
W_{zz}-\left(2m+{\alpha}{}\right)W_{z}+
\left(m^2+{m\,\alpha}{}+{b-C_0}{}\right)W-\frac{1}{2}e^{-m\,z}\,W^2=0.
\end{gathered}\end{equation}

Supposing
\begin{equation}\begin{gathered}
\label{KdVB.11} W(z)=f(y),\qquad y=\varphi(z),
\end{gathered}\end{equation}
we have
\begin{equation}\begin{gathered}
\label{KdVB.12}W_{z}=f_y\,\frac{dy}{d z}, \qquad
W_{zz}=f_{yy}\,\left(\frac{dy}{dz}\right)^2+f_y\,\frac{d^2 y}{d z^2}.
\end{gathered}\end{equation}

Substituting \eqref{KdVB.11} and \eqref{KdVB.12} into Eq.
\eqref{KdVB.10} we obtain the equation
\begin{equation}\begin{gathered}
\label{KdVB.13}f_{yy}\,\left(\frac{dy}{dz}\right)^2-
\frac{1}{2\,}\,e^{-m\,z}\,f^2+f_y\,\left(\frac{d^2y}{dz^2}-
(2\,m+{\alpha}{}\,)\frac{d y}{d z}\right)+\\
\\
+\left(m^2+{\alpha}{}+{b-C_0}{}\right)\,f=0.
\end{gathered}\end{equation}

Assuming in Eq.\eqref{KdVB.13}
\begin{equation}\begin{gathered}
\label{KdVB.14}\left(\frac{dy}{d z}\right)^2=\frac{1}{12\,}\,e^{-m\,z},
\end{gathered}\end{equation}

\begin{equation}\begin{gathered}
\label{KdVB.15}\frac{d^2 y}{d z^2}=\left(2\,m+{\alpha}{}\right)\,\frac{dy}{d z},
\end{gathered}\end{equation}

\begin{equation}\begin{gathered}
\label{KdVB.16}m^2+m\,{\alpha}{}+{b-C_0}{}=0,
\end{gathered}\end{equation}
we have equation
\begin{equation}\begin{gathered}
\label{KdVB.17}f_{yy}=6\,f^2.
\end{gathered}\end{equation}

Multiplying Eq.\eqref{KdVB.17} by $f_y$ and integrating with respect
to $y$, we have
\begin{equation}\begin{gathered}
\label{KdVB.17a}f_{y}^2=4\,f^2-C_2,
\end{gathered}\end{equation}
where $C_2$ is an arbitrary constant. The solution of
Eq.\eqref{KdVB.17a} is found by means of the integral
\begin{equation}\begin{gathered}
\label{KdVB.17b}\int{\frac{df}{\sqrt{4\,f^3-C_2}}}=y
\end{gathered}\end{equation}
and is expressed via the Weierstrass function
\begin{equation}\begin{gathered}
\label{KdVB.17c}f(y)=\wp{(y+C_3,\,0,\,C_2)}
\end{gathered}\end{equation}
with invariants $g_2=0$ and $g_3=C_2$, $C_3$ is an arbitrary
constant.

From Eq. \eqref{KdVB.14} we find $y(z)$ in the form
\begin{equation}\begin{gathered}
\label{KdVB.18}y(z)=C_4-\frac{1}{m\,\sqrt{3\,}}\,e^{-m\,z/2},
\end{gathered}\end{equation}
where $C_4$ is an arbitrary constant. Using $y(z)$  we obtain from
Eqs. \eqref{KdVB.15} and \eqref{KdVB.16} values of $m$ and $b$
\begin{equation}\begin{gathered}
\label{KdVB.19}m=-\frac{2\alpha}{5\,}, \qquad
b= C_0+\frac{6\,\alpha^2}{25\,}.
\end{gathered}\end{equation}

Using this value of $b$ we obtain $C_1$ from Eq.\eqref{KdVB.6} in
the form
\begin{equation}\begin{gathered}
\label{KdVB.20}C_1=\frac{C_0^2}{2}-\frac{18\,\alpha^4}{625\,}.
\end{gathered}\end{equation}

Taking \eqref{KdVB.10a}, \eqref{KdVB.8}, \eqref{KdVB.17},
\eqref{KdVB.18} and \eqref{KdVB.20} into account we have the general
solution of Eq. \eqref{KdVB2} in the form
\begin{equation}\begin{gathered}
\label{KdVB.21}w(z)=C_{0}^{(j)}+\frac{6\alpha^2}{25}-
\exp{\left\{\frac{2\alpha\,z}{5\,}\right\}}\,
\wp{\left(C_3-\frac{5}{\alpha\sqrt{12}}\,
\exp{\left\{\frac{\alpha\,z}{5\,}\right\},\,0,\,C_2}\right)},\\
\\
z=x-C_{0}^{(j)}\,t, \qquad (j=1,2), \qquad C_{0}^{(1,2)}=
\sqrt{2\,C_1+\frac{36\,\alpha^4}{625\,}}.
\end{gathered}\end{equation}

The solitary traveling wave solutions of the KdV-Burgers equation
can be obtained from solution \eqref{KdVB.21} in the case $C_2=0$.
As this takes place the solution  of Eq. \eqref{KdVB.17a} takes the
form
\begin{equation}\begin{gathered}
\label{KdVB.22}f(y)=\frac{1}{(C_{2}\pm y)^2}.
\end{gathered}\end{equation}
Using the solutions \eqref{KdVB.21} and \eqref{KdVB.22} we obtain
the solitary traveling wave solutions in the form
\begin{equation}\begin{gathered}
\label{KdVB.23}w(z)=C_0+\frac{6\,\alpha^2}{25\,
}-\frac{\exp{\left\{\frac{2\,\alpha\,z}{5\,}\right\}}}
{\left(C_3\pm\frac{5\,}{\alpha\,\sqrt{12\,}}
\exp{\left\{\frac{\alpha\,z}{5\,}\right\}}\right)^2}.
\end{gathered}\end{equation}

The solution \eqref{KdVB.23} can be transformed into the ordinary form
\begin{equation}\begin{gathered}
\label{KdVB.24}w(z)=C_0+\frac{6\,\alpha^2}{25\,
}-\frac{12\,\alpha^2} {25\,\,\left(1\pm
C_8\,\exp{\left\{-\frac{\alpha\,z}{5\,}\right\}}\right)^2}.
\end{gathered}\end{equation}

At $C_1=0$  for the solutions \eqref{KdVB.21} and
\eqref{KdVB.24} we have $C_0$ and $z$ in the form
\begin{equation}\begin{gathered}
\label{KdVB.25}C_0^{(1,2)}=\pm\,\frac{6\,\alpha^2}{25\,},\qquad
z=x\mp\,\frac{6\,\alpha^2}{25\,}\,\,t.
\end{gathered}\end{equation}

The solitary wave solutions \eqref{KdVB.24} were found  in \cite{Kudr_88, Kudr_90a} by means of the singular manifold method.

\section{Traveling wave solutions of the Kuramoto - Sivashinsky equation}

The Kuramoto - Sivashinsky equation is one of the popular nonlinear evolution equations which is used for the description of different physical processes. This equation can be written as \cite{Kuramoto, Sivashinsky, Benney, Topper, Shkadov, Cohen}
\begin{equation}\label{KS1}
u_t+\,u\,u_x+\alpha\,u_{xx}+\beta\,u_{xxx}+\gamma\,u_{xxxx}=0
\end{equation}
Using the traveling wave $u(x,t)=w(z)$, where $z=\,x-C_0\,t$ we obtain the nonlinear ordinary differential equation of the third order
\begin{equation}\label{KS0}
\gamma\,w_{zzz}+\beta\,w_{zz}+\alpha\,w_z-\,\frac{w^2}{2}-C_0\,w-C_1=0.
\end{equation}
The general solution has the third order and the Laurent series for the solution of Eq.\eqref{KS0} can be written as
\begin{equation}\begin{gathered}
\label{Ser3a} w(z)=\,{\frac {120\,\gamma}{{(z-z_0)}^{3}}}-\,{\frac {15\,\beta}{{(z-z_0)}^{2}}}+\left({\frac {60\,\alpha}{19}}\,
 -\,{\frac {{15\,\beta}^{2}}{76\,\gamma}}\right) \frac{1}{(z-z_0)}+\\
 \\
 +{C_0}+\,{\frac {7\,\alpha\,\beta}{76\,\gamma}}-{\frac {13}{608}}+\left( {}\,{\frac {87\alpha\,{\beta}^{2}}{{5776\,\gamma}^{2}}}-{
}\,{\frac {{131\,\beta}^{4}}{{46208\,\gamma}^{3}}}-{\frac {}{}}
\,{\frac {{11\,\alpha}^{2}}{722\,\gamma}} \right) (z-z_0)+\\
\\
+\left( {}\,{\frac {{15\,\beta}^{3}\alpha}{{5776\,\gamma}^{3}}}-{
}\,{\frac {{9\,\beta}^{5}}{{23104\,\gamma}^{4}}}-{}\,
{\frac {25\,\beta\,{\alpha}^{2}}{{5776\,\gamma}^{2}}} \right) {(z-z_0)}^{2}
+\\
\\
+\left({\frac {{
{C_0}}^{2}+2\,C_1}{252\gamma}}\right)(z-z_0)^3+\ldots
\end{gathered}\end{equation}

From the Laurent series \eqref{Ser3a} we can see that the solution of the Kuramoto - Sivashinsky equation \eqref{KS0} has the pole of the third order. Using the Laurent series \eqref{Ser3} for the Riccati equation we can write
\begin{equation}\label{KS2a}
w(z)=A_0+A_1\,w_1(z)+A_2\,w_1(z)^2+A_3\,w_1(z)^3.
\end{equation}
Substituting the Laurent series \eqref{Ser3a} for $w(z)$ and the Laurent series $w_1(z)$ for the Riccati equation into Eq.\eqref{KS2a} we have
\begin{equation}
\begin{gathered}\label{KS2b}
A_3=120\,\gamma, \quad A_2=-15\,\beta,\quad  A_1=-{\frac {15}{76}}\,{\frac {{\beta}^{2}}{\gamma}}+{\frac {60}{19}}\,\alpha-
120\,\lambda\,\gamma,\\
\\
A_0=-{\frac {13}{608}}\,{\frac {{\beta}^{3}}{{\gamma}^{2}}}+10\,\beta\,\lambda+{
\frac {7}{76}}\,{\frac {\alpha\,\beta}{\gamma}}+{C_0}
\end{gathered}\end{equation}
We have two values for $\lambda$
\begin{equation}
\label{KS2c}\begin{gathered}
\lambda_{1,2}=\pm{\frac {5}{76}}\,{\frac {\alpha}{\gamma}}\mp{\frac {3}{152}}\,{\frac {{\beta}
^{2}}{{\gamma}^{2}}},
\end{gathered}\end{equation}
and two values for $C_1$
\begin{equation}
\begin{gathered}\label{Parameters0}
C_{1}^{(1)}={\frac {50671}{14047232}}\,{\frac {{\beta}^{6}}{{\gamma}^{4}}}-{\frac {6753
}{219488}}\,{\frac {{\beta}^{4}\alpha}{{\gamma}^{3}}}+{\frac {3813}{54872}}
\,{\frac {{\beta}^{2}{\alpha}^{2}}{{\gamma}^{2}}}-{\frac {72}{6859}}\,{
\frac {{\alpha}^{3}}{\gamma}}-\frac{{{C_0}}^{2}}{2},\,\\
\\
C_{1}^{(2)}={\frac {119791}{14047232}}\,{\frac {{\beta}^{6}}{{\gamma}^{4}}}-{\frac {
17553}{219488}}\,{\frac {{\beta}^{4}\alpha}{{\gamma}^{3}}}+{\frac {12813}{
54872}}\,{\frac {{\beta}^{2}{\alpha}^{2}}{{\gamma}^{2}}}-{\frac {1322}{6859
}}\,{\frac {{\alpha}^{3}}{\gamma}}-\frac{{{C_0}}^{2}}{2}.
\end{gathered}\end{equation}
We also obtain seven relations  for parameters $\alpha$, $\beta$ and $\gamma$ of Eq.\eqref{KS0}
\begin{equation}\label{Parameters1}
\frac{\beta}{\sqrt{\alpha\,\gamma}}=0, \quad\pm\frac{12}{\sqrt{47}}, \quad\pm\frac{16}{\sqrt{73}},\quad \pm 4.
\end{equation}

Solitary wave solutions of the Kuramoto - Sivashinsky equations are expressed by formula
\begin{equation}\begin{gathered}
\label{Parameters2}w(z)={C_0}+ 10\,\beta\,\lambda-{\frac {13}{608}}\,{\frac {{\beta}^{3}}{{\gamma}^{2}}
}+{\frac {7}{76}}\,{\frac {\alpha\,\beta}{\gamma}}+\\
\\
+ \left({\frac {60}{19}}\,\alpha-{\frac {15}{76}
}\,{\frac {{\beta}^{2}}{\gamma}}-120\,\gamma\lambda \right)
\sqrt {\lambda}\tanh \left( \sqrt {\lambda}z \right) -\\
\\
-
15\,\beta\,\lambda \, \tanh^{2}
\left( \sqrt {\lambda}z \right) +120\,\gamma{\lambda}^{3/2} \tanh^{3}\left(
\sqrt {\lambda}z \right).
\end{gathered}\end{equation}
Values $\lambda$ and constants $C_1$ are determined by formulae Eq.\eqref{KS2c} and Eq.\eqref{Parameters0}. As this takes place we have to take the corresponding relations \eqref{Parameters1} into account for parameters $\alpha$, $\beta$ and $\gamma$ of Eq.\eqref{KS0}. Exact solutions of Eq.\eqref{KS0} were first found at $\beta=0$ in \cite{Kuramoto} and in the case $\beta\neq0$ in works \cite{Kudr_88, Kudr_90a, Kudr_90, Kudr_91, Kudr_96, Kudr_07,  Berloff, Fu, Zhu}.
In the case $\beta^2=16 \,\alpha\,\gamma$ Eq.\eqref{KS0} has a solution expressed by the Jacobi elliptic function \cite{Kudr_90}. It has been proved  by Eremenko in papers \cite{Eremenko_2005, Eremenko_2006}. We can not find other solitary and periodic wave solutions of Eq.\eqref{KS0}.

Assuming $\alpha=\beta=0$ we have the partial case of the Kuramoto - Sivashinsky equation in the form
\begin{equation}\label{KS4}
u_t+u\,u_x+\gamma\,u_{xxxx}=0.
\end{equation}

Taking the traveling wave $w(z)$, $z=x-C_0\,t$  into account we can reduce Eq.\eqref{KS4} to the following
\begin{equation}\label{KS4a}
\gamma\,w_{zzz}+\frac12\,w^2-C_0\,w-C_1=0.
\end{equation}

From the Laurent series \eqref{Ser3a} we have the following expansion for solution of Eq.\eqref{KS4a}
\begin{equation}\label{KS5}
w(z)={\frac {120\,\gamma}{{(z-z_0)}^{3}}}+{C_0}+\,{\frac { \left( {
{C_0}}^{2}+2\,{C_1} \right) {(z-z_0)}^{3}}{252\,\gamma}}+\ldots
\end{equation}
However, using the Laurent series \eqref{KS5} for solution of Eq.\eqref{KS4a} we can find the only rational solution at $C_1=-\frac{C_{0}^{2}}{2}$
\begin{equation}\label{KS6}
w(z)={\frac {120\,\gamma}{{(z-z_0)}^{3}}}+{C_0}.
\end{equation}
Comparison of the Laurent series for solutions of Eq.\eqref{BE2} and Eq.\eqref{KS4a} does not allow us to find any other solutions of Eq.\eqref{KS4a}.

\section{Traveling wave solutions of the Kawahara equation}

The Kawahara equation takes the form
\begin{equation}\label{KE1}
u_t+\,u\,u_x+\beta\,u_{xxx}=\delta\,u_{xxxxx}
\end{equation}
Taking the traveling wave into account we obtain the nonlinear ordinary differential equation \cite{Assas_2009}
\begin{equation}\label{KE0}
\delta\,w_{zzzz}-\beta\,w_{zz}-\,\frac{w^2}{2}-C_0\,w-C_1=0.
\end{equation}

Multiplying Eq.\eqref{KE0} on $w_z$ we have the first integral in the form
\begin{equation}\label{KE0a}
\delta\,w_{zzz}\,w_{z}-\frac12\,\delta\,w_{zz}^2-\frac12\,\beta\,w_z^2-
\,\frac{w^3}{6}-\frac12\,C_0\,w^2-C_1\,w+C_2=0.
\end{equation}

The Laurent series for the fourth order nonlinear differential equation can be written as

\begin{equation}\begin{gathered}
\label{Ser3d} w(z)={\frac {1680\,\delta}{{(z-z_0)}^{4}}}-{}\,{\frac {280\,\beta}{13{(z-z_0)}^
{2}}}-{C_0}-\,{\frac {{31\,\beta}^{2}}{507\,\delta}}-{
}\,{\frac {{31\,\beta}^{3}}{{39546\,\delta}^{2}}}\,{(z-z_0)}^{2}+\\
\\
+
\left( {\frac {{{C_0}}^{2}}{3312\,\delta
}}\,{}-{\frac {{C_1}}{1656\,\delta}}\,{}-{\frac {2945\,{\beta}^{4}}{283782096\,{\delta}^{3}}}\,{
\frac {}{}} \right) {(z-z_0)}^{4}+\ldots
\end{gathered}\end{equation}

Equation does not pass the Painlev\'e test and the Cauchy problem can not be solved by the inverse scattering transform but one can obtain some exact solutions of this equation. There is a solution expressed via the Jacobi elliptic function \cite{Kudr_90a}. The solitary wave solutions can be found by many methods.

Let us demonstrate that we can find the solitary wave solutions of the Kawahara equation using the Laurent series. Assuming $w(z)$ in the form
\begin{equation}\begin{gathered}
\label{Eq2}w(z)=A_0+A_1\,w_1(z) +A_{{2}}w_1(z)^2
 +A_{{3}} \,w_1(z)^3 + A_{{4}}\,w_1(z)^4,
  \end{gathered}\end{equation}
where $w_1(z)$ is the Laurent series of the Riccati equation \eqref{BE2}, and substituting \eqref{Ser3} and \eqref{Ser3d} into Eq.\eqref{Eq2} we have
\begin{equation}\begin{gathered}
\label{Eq3}A_{{4}}=\,{ {1680\,\delta}}, \quad A_{3}=0,\qquad A_{{2}}=-{\frac {280}{13}}\,\beta-2240\,\delta\,\lambda,\qquad A_{1}=0, \\
\\
A_{{0}}={C_0}+{\frac {1568}{3}}\,\delta\,{\lambda}^{2}-{\frac {31}{507}}\,{\frac
{{\beta}^{2}}{\delta}}+{\frac {560}{39}}\,\beta\,\lambda.
\end{gathered}\end{equation}
We also have three formulae for $\lambda$
\begin{equation}\begin{gathered}\label{par1}
\lambda_1={\frac {1}{52}}\,{\frac {\beta}{\delta}},\quad \lambda_{2,3}=
{\frac {\beta}{52\,\delta}}\,{ { \left( -{ \frac{31}{20}}\pm{\frac {3}{20}}\,i
\sqrt {31} \right) }{}}
 \end{gathered}\end{equation}
and for constant $C_1$
\begin{equation}\begin{gathered}\label{par1a}
C_1={\frac {2945}{171366}}\,{\frac {{\beta}^{4}}{{\delta}^{2}}}-\frac{{{
C_0}}^{2}}{2}+{\frac {14720}{39}}\,\delta\,{
\lambda}^{3}\beta\,+20608\,{\delta}^{2}{\lambda}^{4}.
 \end{gathered}\end{equation}

As a result we obtain the solitary wave solutions in the form
\begin{equation}\begin{gathered}
\label{Eq4}u(x,t)={C_0}-{\frac {31}{507}}\,{\frac {{\beta}^{2}}{\delta}}
+{\frac {1568
}{3}}\,\delta\,{\lambda}^{2}+{\frac {560}{39}}\,\lambda\beta+ \\
\\
-\left(2240\,\delta
\,\lambda+{\frac {280}{13}}\,\beta \right)\lambda \tanh^{2} \left( \sqrt {\lambda}z
 \right)+1680\,\delta\,{\lambda}^{2} \tanh^{4} \left(
\sqrt {\lambda}z \right),
\\
\\
\quad z=x-C_0\,t,
  \end{gathered}\end{equation}
where parameters $\lambda$ and $C_1$ are determined by formulae \eqref{par1} and \eqref{par1a}.
Instead of the function $w_1(z)$ we can take in \eqref{Eq2} the functions  $w_2(z)$, $\ldots, w_9(z)$, but all these solutions of the Kawahara equation are the same.

Recently Assas \cite{Assas_2009} considered the Kawahara equation using the traveling wave
$z=k\,x-C_0\,t$ and  tried to find new exact solutions of the nonlinear ordinary differential equation
\begin{equation}\label{Eq1a}
\gamma\,k^5\,U_{zzzzz}-k^3\,\beta\,U_{zzz}-\alpha\,k\,U\,U_{z}+C_0\,U_{z}=0.
\end{equation}

He believed that he found  few exact solutions of Eq.\eqref{Eq1a} but this is not the case.  It is obvious that the author \cite{Assas_2009} can not obtain any new  solutions of Eq.\eqref{Eq1a} using the Exp-function method. The reason is that the general solution of Eq.\eqref{Eq1a} has the pole of the fourth order.  This fact  has an important bearing on the choose of the expressions in the Exp-function method. However the author did not take this result into account. To look for exact solutions of the Kawahara equation the author \cite{Assas_2009} must have taken at least five terms in denominator and numerator of the ansatz in the Exp-function method. We can see that all constructions by author (formulae (15), (21) and (27) of work \cite{Assas_2009}) have poles of the second and the third order.

Consequently we expected that Assas could not find any solutions of the Kawahara equation. Just to be on the safe side we have checked all solutions by Assas \cite{Assas_2009} and obtained that all solutions \cite{Assas_2009} do not satisfy the Kawahara equation.

\section{Partial case of the Kawahara equation}

Consider the partial case of the Kawahara equation
\begin{equation}\label{Kawm1}
u_t+\,u\,u_x=\delta\,u_{xxxxx}
\end{equation}
Taking the traveling wave $u(x,t)=w(z)$, where $z=x-C_0\,t$ into account we have the nonlinear ordinary differential equation of the fourth order \cite{Inan}
\begin{equation}\label{KE00}
\delta\,\,w_{zzzz}-\frac{\,w^2}{2}-C_0\,w-C_1=0.
\end{equation}

The Laurent series for the fourth order nonlinear differential equation is read as

\begin{equation}\begin{gathered}
\label{Ser33a} w(z)={\frac {1680\,\delta\,{}}{{(z-z_0)}^{4}}}-{{{C_0}}}+
 \left( {}\,{\frac {{{C_0}}^{2}}{3312\,\delta\,{}}}-{
}\,{\frac {{C_1}}{1656\,\delta\,{}}} \right) {(z-z_0)}^{4}+\ldots
\end{gathered}\end{equation}

Eq.\eqref{KE00} does not pass the Painlev\'e test. The Cauchy problem can not be solved for this equation by the inverse scattering transform. From the Laurent series \eqref{Ser33a} we can find the rational solution of Eq.\eqref{KE0} at $C_1=-\frac{C_{0}^{2}}{2}$ in the form
\begin{equation}\begin{gathered}
\label{Sol33a} w(z)={\frac {1680\,\delta\,}{{(z-z_0)}^{4}}}-{ {{C_0}}{}}.
\end{gathered}\end{equation}
We can not find any other solutions of Eq.\eqref{KE00} from the comparison of the Laurent series \eqref{Ser33a} for solution of Eq.\eqref{KE00} and the Lorent series of the Riccati equation \eqref{BE0}. We could expect that all "exact solutions" of Eq.\eqref{Kawm1} by Inan and Ugurlu \cite{Inan} are wrong. We have checked these solutions and realized that expressions by Inan and Ugurlu do not satisfy Eq.\eqref{Kawm1}. These expressions are not exact solutions of Eq.\eqref{Kawm1}.

\end{document}